\documentclass[journal]{IEEEtran}

\ifx\pdfoutput\undefined
\usepackage{graphicx}
\else
\usepackage[pdftex]{graphicx}
\fi

\ifx\pdfoutput\undefined
\usepackage[hypertex]{hyperref}
\else
\usepackage[pdftex,hypertexnames=false]{hyperref}
\fi
\begin{document}
%
\title{Tradeoff between Efficiency and Melting for a High-Performance Electromagnetic Rail Gun}
%
%
\author{\authorblockN{William C. McCorkle and Thomas B. Bahder}\\
\authorblockA{{Army Aviation and Missile Research, Development, and Engineering Center,\\ 
Redstone Arsenal, AL  35898 USA}\\
Email: thomas.bahder@us.army.mil}}
%



\maketitle

\begin{abstract}
We estimate the temperature distribution  in the rails of an electromagnetic rail gun (EMG) due to the confinement of the current in a narrow surface layer resulting from the skin effect.  In order to obtain analytic results, we assume a simple geometry for the rails, an electromagnetic skin effect boundary edge that propagates with the  accelerating armature, and a current carrying channel controlled by magnetic field diffusion into the rails.  We compute the temperature distribution in the rails at the time that the armature leaves the rails. For the range of exit velocities, from 1500 m/s to 5000 m/s, we find the highest temperatures are near the gun breech. After a single gun firing, the temperature reaches the melting temperature of the metal rails in a layer of finite thickness near the surface of the rails, for rails made of copper or tantalum.  We plot the thickness of the melt layer as a function of position along the rails.  In all cases, the thickness of the melt layer increases with gun velocity, making damage to the gun rails more likely at higher velocity.   We also calculate the efficiency of the EMG as a function of gun velocity and find that the efficiency increases with increasing velocity, if the length of the gun is sufficiently long.   The thickness of the melted layer also decreases with increasing rail length.  Therefore, there is a tradeoff:  for rails of sufficient length, the gun efficiency increases with increasing velocity but the melted layer thickness in the rails also increases.
\end{abstract}


%
\IEEEpeerreviewmaketitle

\section{Introduction}
%
%
%
%

\PARstart{E}{lectromagnetic} launch systems, such as the railgun, are based on transient phenomena~\cite{14th_Symposium_2008}. During launch, the transient involves the build-up and penetration of a magnetic field into the surrounding metallic material.  The dynamics of magnetic field penetration into the metal rails is described by a well-known diffusion equation~\cite{Knoepfel2000}. The diffusion of the magnetic field leads to a skin effect, where a large current is transported inside a narrow channel.   In a railgun that has a moving conducting armature, the effect is called a velocity skin effect (VSE) and is believed to be one of the major problems in limiting railgun performance~\cite{Knoepfel2000,Young1982}, because it leads to intense Joule heating of the conducting materials, such as rails and armatures. To what extent the VSE effect is responsible for limiting the performance of solid armatures is still the subject of research~\cite{Drobyshevski1999,Stefani2005}.  

Recently, using a new generation of magnetic field sensors, the magnetic field distributions caused by the VSE in the rails have been measured~\cite{Schneider2007,Schneider2009}.  These experimental efforts are even more significant in light of the large investments planned by the navy to  develop EMGs and power sources for nuclear and conventional warships~\cite{Walls1999,mcnab2007}. The motivation for this paper is the large investment planned for EMG technology and the historical lack of understanding of the reasons for the low endurance of the gun rails in service, sometimes limited to one shot at maximum energies before replacement is needed.  

For high-performance EMGs, in order to increase the armature velocity while keeping the length of the rails fixed, the  current pulse during firing must be shorter and have a higher average amplitude, causing a stronger skin effect in the rails, which leads to an increase in Joule heating of the rails.  In this paper, we show that the EMG efficiency is higher at higher velocity, but there is increased  melting of the rails, leading to a tradeoff between efficiency of the EMG and melting of the rails due to Joule heating (during a single firing). We do not discuss gun barrel erosion due to repeated firings~\cite{Ahmad1988,Johnston2005}. We also do not treat the interaction of the high temperature plasma in the contact regions of the EMRG. 

The article is organized as follows. We define our model for EMG heating in Section II.  We consider the problem of rail heating in two steps.  First, in Section~III we assume a current-carrying channel described by a local skin depth $\delta$ (along the rail) that has a simple time dependence due to motion of the armature, which leads to a time and position-dependent current density that causes the Joule heating.  In Section~IV, we use an improved expression for the current density based on the diffusion of the magnetic field into the rails.  In Section V, we give a crude approximation to the armature heating. In Section VI, we discuss the efficiency of the EMG and its dependence on gun velocity and length.   We show that the there is a tradeoff between efficiency and melting the rails. Finally, we present our conclusions in Section VII.

\section{Temperature Distribution in Rails}

The dynamics of an EMG can be properly described in terms of a thermodynamic free energy that expresses the coupling of the  mechanical and electromagnetic degrees of freedom~\cite{LL_continuous_media}.  For example, for the case of an electromagnetic gun that can be described in terms of a lumped circuit model, which has a rotor coil with self inductance $L_R$ carrying current  $I_R$, and a stator coil, rail, and armature circuit with self inductance $L$ with current $I$, the free energy has the form~\cite{bahder_bruno_1998} 
\begin{equation}
F(I_R,I,x, \theta)= F_0(T) + \frac{1}{2} L_R I_R^2 + \frac{1}{2} L(x) I^2 + M(\theta) \,  I_R \, I 
\label{freeEenergy}
\end{equation}
where $L(x)$ depends on the $x$-position of the armature,  
\begin{equation}
L(x) = L_0 \, + \, L^\prime  \,\, x
\label{variableInductance}
\end{equation}
and where $L^\prime = d L_S /dx$, and $L_0$ is the self inductance of the stator, rail and armature circuit when the armature is at $x=0$.   The interaction of the stator and rotor circuits is specified in terms of their  mutual inductance, $M(\theta)$, where $\theta$ is the angle of the rotor coil with respect to the stator coil. The term $F_0(T)$ depends only on temperature.  Derivatives of the free energy with respect to the coordinates, $x$ and $\theta$, give the generalized forces on the system~\cite{LL_continuous_media}.  By Newton's law, the acceleration of the armature, $\ddot x$, is given in terms of the derivative of the free energy with respect to the coordinate, 
\begin{equation}
m\,\ddot x = \left( {\frac{{\partial {\kern 1pt} F}}{{\partial \,x}}} \right)_{I_R , I, \theta } 
\label{free_energy}
\end{equation}
where $m$ is the mass of the armature (including the payload, comprising the launch package), leading to the well-known dynamical equation
\begin{equation}
m\,\ddot x = \frac{1}{2}{\kern 1pt} L'{\kern 1pt} I^2 
\label{acceleration}
\end{equation}
Equation~(\ref{acceleration}) shows that the acceleration is directly proportional to the square of the instantaneous current, $I$. As a first approximation, we assume that the current in the rails is constant during a shot, $I(t)=I_o$.  If the current is constant, and the armature starts at $t=0$ at $x=0$, and moves to the end of the gun rails at $x= \ell$ at time $t=t_f$, then we have the following relations between armature mass $m$, gun length $\ell$, self inductance per unit length of armature travel $
L'$, and mass velocity $v$,
\begin{equation}
\begin{array}{l}
I_o  = \left( {\frac{{m\,}}{{L'\, \ell }}} \right)^{1/2} \,v \\ 
t_f  = \frac{2}{{I_o }}\left( {\frac{{m\, \ell }}{{L'}}} \right)^{1/2}  \\ 
t_o (x) = \frac{2}{{I_o }}\left( {\frac{{x{\kern 1pt} m}}{{L'}}} \right)^{1/2}  \\ 
 \end{array} \label{constantCurrent}
\end{equation}
The function $t_o (x)$ gives the time at which the mass $m$ is at position $x$ along the rails, see section III.

In order to calculate the Joule heating in the rails,  a lumped circuit model is not sufficiently detailed. Instead, we must use a more detailed model, where the free energy is expressed in terms of electromagnetic fields with a spatial distribution.  The problem is complicated due to the coupling of the electromagnetic and mechanical degrees of freedom. The problem is further complicated by the fact that the dynamics of an EMG shot is a transient effect in time.  Consequently, when Long~\cite{Long1986} and  Nearing and Huerta~\cite{Nearing1999}  computed  current density (and the heating in   EMG rails) they assumed that the mechanical and electromagnetic degrees of freedom are decoupled.  Furthermore, they assumed the armature was moving at a constant speed so the problem became translationally invariant in time, thereby avoiding the complexities associated with the initial conditions and the resulting transient effects. In particular, it is the transient nature of the EMG shot that gives rise to a dependence of EMG performance on rail length, see for example our Eq.~(\ref{temprise3}) and~(\ref{efficiency}) for dependence of rail temperature rise and EMG efficiency, respectively, on rail length $\ell$.   Using the above stated  assumptions, Long~\cite{Long1986} and  Nearing and Huerta~\cite{Nearing1999}  solved for the complicated distribution of the current density using a simplified geometric model of the rails and armature~\cite{Powell_ARLTR_2008}.  

In this paper, we use a simpler approach that addresses the transient nature of the EMG shot and allows us to get approximate analytic results for the temperature distribution in the rails and armature of an EMG.   We {\it model} the skin effect in the rails, which limits the channel through which current can flow. For a given total current in the rails, a narrower channel (smaller skin depth) leads to a higher current density and results in greater Joule heating of the rails.  In contrast to Nearing and Huerta~\cite{Nearing1999}, our approach allows us to discuss transient effects dependent on the length of EMG rails and how they impact EMG performance, see  for example our Eq.~(\ref{temprise3}) and~(\ref{efficiency}).   We use the same 2-dimensional simplified geometry for the rails and armature as Nearing and Huerta. However, we assume that the conducting armature is arbitrarily thin, which allows us to get simple results.  Corrections to such an assumption are expected to be of order $O(a/ \ell)$, where $a$ is the armature width and $\ell$ is the length of the rails.   We take the coordinate $x$ running down the length of the rails, and we assume that the top rail occupies $y>0$ and the bottom rail occupies $y<-b$, where $b$ is the rail separation. We take the length of the gun from breech to muzzle to be $\ell$. We assume that the rails have an arbitrary large thickness $w$ in the $z$-direction, see Figure~1.  When an EMG is fired, the current flows down one rail, through the conducting armature, and up the other rail to complete the electrical circuit. The transient response leads to a complicated distribution of eddy currents in the rails and the armature.  We assume that we can describe this effect by the local skin effect that depends on position and time. 

We take the temperature rise due to the Joule heating to be 
\begin{equation}
T(x,y,t_f ) - T_0  = \frac{{1\,}}{{\rho \,\,{\kern 1pt} C}}\int\limits_0^{t_f } {\,\frac{{J^2 (x,y,t)}}{\sigma }{\kern 1pt} \,dt} 
\label{heating}
\end{equation}
where  $T(x,y,t_f )$ is the temperature at position $(x,y)$ at time $t_f$, when the armature leaves the rails, assuming the current starts at $t=0$.  The quantity $T_0 =T(x,y,0)$ is the initial temperature at position $(x,y)$ at time $t=0$ before the shot, $C$ is the specific heat (assumed constant up to the melting point) of the metal rails, and $\rho$  is the density of the metal rail.  We  assume the electric field $E(x,y,t)$ is linearly related to the current density, $J(x,y,t)= \sigma \, E(x,y,t)$, where $\sigma$ is the electrical conductivity that is independent of temperature up to the melting point.  In Eq.~(\ref{heating}), we have neglected the heat of melting, so the temperature rise is only valid up to the melting point of the metal rail.  If latent heat of melting $L_Q$ is included, then the term $-L_Q/C$ must be added to the right side of Eq.~(\ref{heating}).  This additional term subtracts from the temperature rise that may be expected when additional Joule heat is created beyond what is required to reach the melting temperature.    For Cu or Ta, this term is significant, with value $L_Q/C=465\,^{\circ}{\rm C}$  and  $1141 \,^{\circ}{\rm C}$, respectively.  Below, we do not consider the temperature rise above  the melting point of the metal rails.

\section{Current Channel with Constant Current in Rails}

\begin{figure}[tbp] 
  \centering
  \includegraphics[width=2.5in]{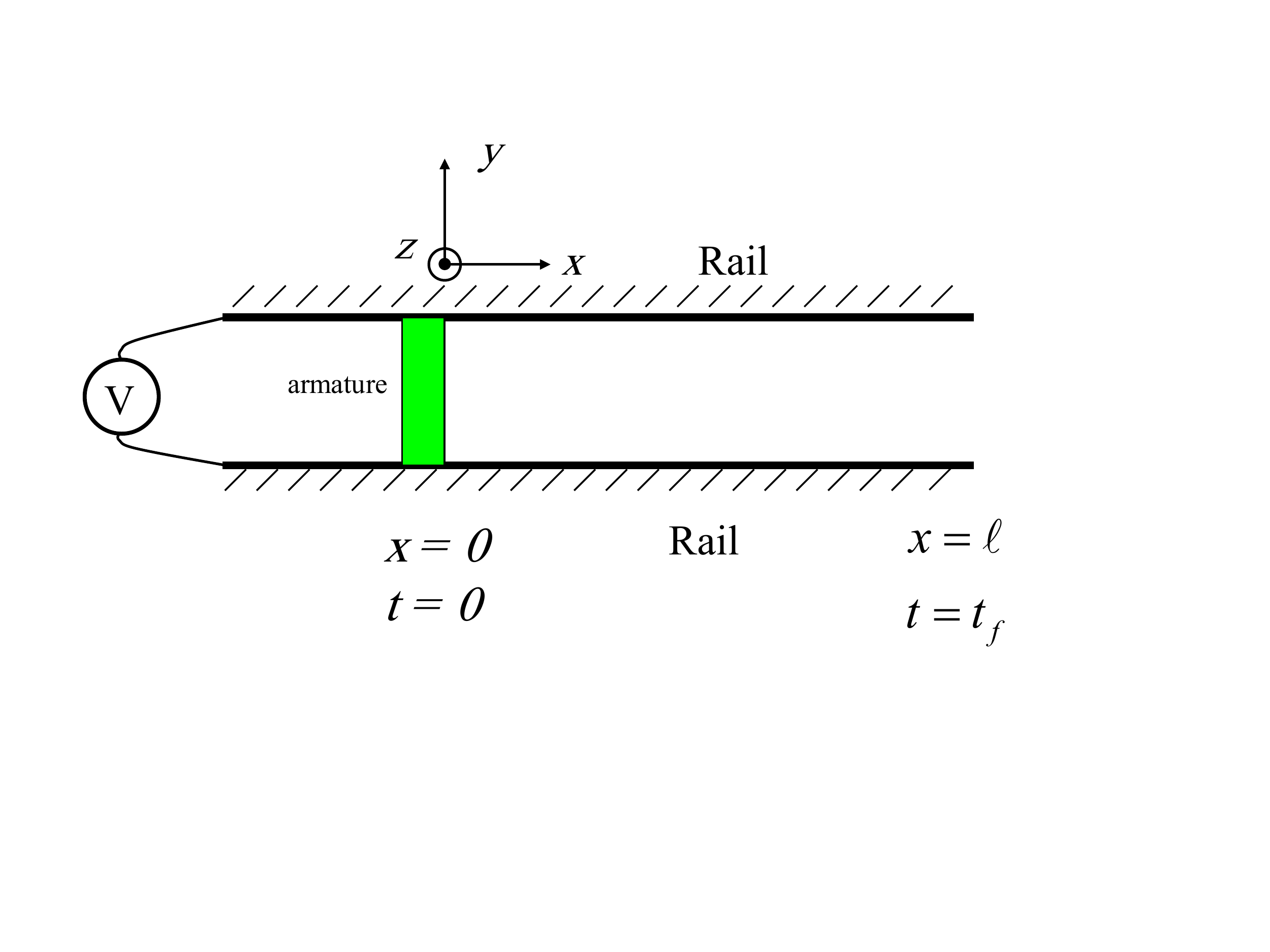}
  \caption{The rails and conducting armature are shown for the electromagnetic gun.  At time $t=0$ the leading edge of the armature is at $x=0$. The trailing edge of the armature leaves the rails at time $t=t_f$.}
  \label{fig:EMRG_Figure1}
\end{figure}

The moving armature carries all the current of the rails.  As the armature moves, it exposes a new plane region on the rail that carries current. Due to the diffusive nature of the magnetic field ${\bf H}$, and the relation 
\begin{equation}
{\bf J} = \textrm{curl} \,\, {\bf H}
\label{MaxwellCurrentDensity}
\end{equation}
current does not flow uniformly in the rail~\cite{LL_continuous_media,Knoepfel1970}, instead the current flows through a layer of thickness $\delta$ that increases with time $t$, starting with $\delta=0$ as the leading edge of the armature passes a point on the rail.  Assuming a plane geometry for the rail,  we can approximate the time dependence of the skin depth to be~\cite{Knoepfel1970}    
\begin{equation}
\delta(t)  = \left( {\frac{{4 \, t}}{{\mu  \, \sigma }}} \right)^{1/2} 
\label{skindepth}
\end{equation}
where $\mu$ is the magnetic permeability and $\sigma$ is the electrical conductivity, and $t$ is the time elapsed since the armature has passed a given element on the rail.  The exact factor inside the square root (here we take it to be 4) is somewhat arbitrary in defining a skin depth. (In Section IV we remove this arbitrariness by using the solution of the  time-dependent magnetic field diffusion equation to compute the current density distribution.) As stated in the second to last paragraph in Section II, we assume an idealized armature that is arbitrarily thin in the $x$-direction and we assume that the skin depth $\delta$ in the rail starts at zero thickness at the position where the armature contacts the rails.  As remarked above,  this drastic assumption is expected to have corrections of order $O(a / \ell)$ where $a$  is the width of the armature and $\ell$ is the length of the rail.  This assumption allows us to obtain analytic results and see the dependence on a number of parameters. As a start, we model the conducting channel by assuming  that the current density in the rail inside the skin depth $\delta$ is a function of $x$ and $t$ but not $y$, and that the current density is zero outside the skin depth $\delta$, see Figure 1~\cite{explain}. Furthermore, we assume that the skin depth has zero thickness on the leading edge of the armature, since all the current in the rails has to flow through the armature. Assuming a current carrying channel of finite width $\delta$, at time $t$ and position $x$ along the rail, the $y$ coordinate of the boundary of the current carrying channel is 
\begin{equation}
y_c(x,t)  = \left\{ {\begin{array}{*{20}c}
   {\delta (t - t_o (x))\,,\quad t > t_o (x)}  \\
   {0\,,\quad \quad \quad \quad \quad t \le t_o (x)}  \\
\end{array}} \right.  \label{channel}
\end{equation}
where $t_o(x)$ is the time the leading edge of the armature passes the position $x$. In other words,  
\begin{equation}
\frac{{d\,t_o (x)}}{{d\,x}} = \frac{1}{{v(x)}}
\end{equation}
where $v(x)$ is the velocity of the armature when it is at point $x$, see Figure~(\ref{fig:EMRG_Figure1}).
 
Consider an element of volume $dV = w dx \, dy$ at position $(x,y)$. Define  $\tau(x,y)$ as the time at which the boundary of the current carrying channel intersects this volume element.  From Eq.~(\ref{channel}) for the time-dependent boundary of the current carrying channel, we find
\begin{equation}
\tau (x,y) = t_o (x) + \frac{1}{4}\mu \sigma y^2 
\label{tau}
\end{equation}
For time $0<t\le \tau(x,y)$ there is no current flowing through this volume element.  At time $t=\tau(x,y)$, the boundary of the current carrying channel intersects the volume element at position $(x,y)$ and Joule heating starts.  See Figures 2, 3 and 4. Finally, at time $t=t_f$ the armature leaves the rails, the circuit is broken, and there is no more Joule heating of the element $dV$.

The current density in the rail can be written down by considering three domain regions, see Figure~5.  At position $(x,y)$ and time $t$, we take the current density to have the form
\begin{equation}
\begin{array}{l}
 J(x,y,t) =  \\ 
 \quad \left\{ {\begin{array}{*{20}c}
   {\frac{{I(t)}}{{w\,\delta (t - t_o (x))}},} & {\quad x < x_o (t)\quad {\rm{and}}\quad y \le \delta (t - t_o (x))\quad }  \\
   {0,} & {x < x_o (t)\quad {\rm{and}}\quad y > \delta (t - t_o (x))}  \\
   0 & {x > x_o (t)}  \\
\end{array}} \right. \\ 
 \end{array}
\label{JcurrentDEF_z}
\end{equation}
where  $x_o(t)$ is the function that gives the $x$ coordinate of the armature at time $t$, and $\delta(t)$ is given by Eq.(\ref{skindepth}).

\begin{figure}[tbp] 
  \centering
  \includegraphics[width=2.5in]{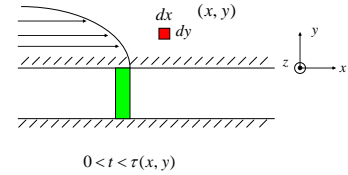}
  \caption{At time $t < \tau $ the skin depth boundary has not yet reached the element of volume $dV= w \, dx \, dy$ at position $(x,y)$. }
  \label{fig:EMRG_Figure2}
\end{figure}

\begin{figure}[tbp] 
  \centering
  \includegraphics[width=2.5in]{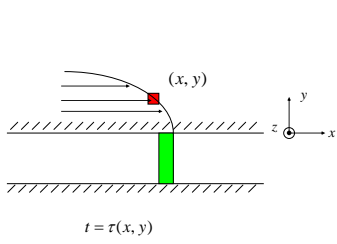}
  \caption{At time $t = \tau$ the skin depth boundary overlaps the element of volume $dV= w \, dx \, dy$ at position $(x,y)$.  At this time, current starts to flow in the element $dV$ and the temperature starts to rise due to Joule heating.}
  \label{fig:EMRG_Figure3}
\end{figure}

\begin{figure}[tbp] 
  \centering
  \includegraphics[width=2.5in]{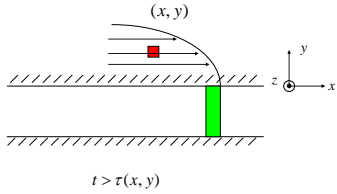}
  \caption{After the skin depth boundary passes the element of volume $dV$, heating of the element continues until the armature exits the rails at the time $t_f$. For time $t> t_f$, we assume that no energy is  input into the element $d V$. }
  \label{fig:EMRG_Figure4}
\end{figure}

\begin{figure}[tbp] 
  \centering
  \includegraphics[width=2.5in]{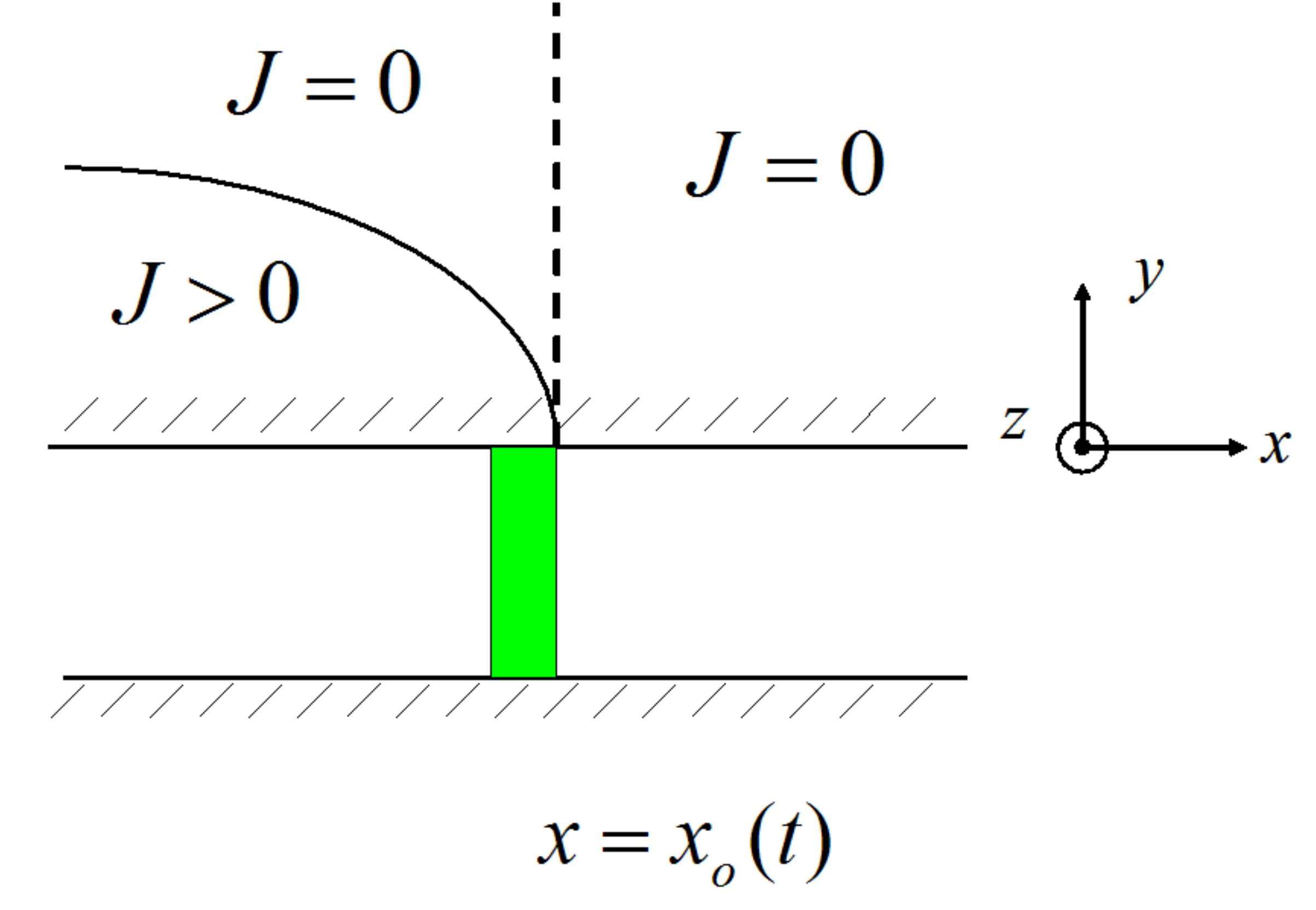}
  \caption{The three domain regions are shown for the current density.  The non-zero current density is assumed to be inside the skin depth for $x < x_o(t)$ where $x_o(t)$ gives the $x$-coordinate of the armature at time $t$.}
  \label{fig:EMRG_Figure5}
\end{figure}

Using Eq.~(\ref{heating}) and assuming a constant current in the armature, $I(t)=I_0$,  from Eq.~(\ref{heating}) and (\ref{JcurrentDEF_z}) we obtain 
\begin{equation}
\begin{array}{l}
 T(x,y,t_f ) - T_0  =  \\ 
 \;\frac{{\mu \,I_0 ^2 }}{{4w^2 \rho C}}\;\left\{ {\begin{array}{*{20}c}
   {Log\left( {\frac{{t_f  - t_o (x)}}{{\tau (x,y) - t_o (x)}}} \right),} & {y \le \left[ {\frac{4}{{\mu \sigma }}\left( {t_f  - t_o (x)} \right)} \right]^{1/2} }  \\
   {0,} & {y > \left[ {\frac{4}{{\mu \sigma }}\left( {t_f  - t_o (x)} \right)} \right]^{1/2} }  \\
\end{array}} \right. \\ 
 \end{array}
\label{tempRiseDistributionConstantCurrent_1}
\end{equation}

Equation~(\ref{tempRiseDistributionConstantCurrent_1}) gives the temperature rise, $T(x,y,t_f ) - T_0$, of an element of volume at position $(x,y)$ at time $t_f$ at which the armature leaves the rails, for a constant gun current over the time interval $0 < t < t_f$.  For  realistic time-dependent currents, see Figure 3 in McCorkle~\cite{McCorkle}. 

Using a lump circuit model to describe the constant acceleration $\ddot{x}=a$ of the armature for a constant current $I_0$,  we can write the position of the armature as a function of time as $x= \frac{1}{2}a t^2 $.  The time at which the armature leaves the rails, $t_f$, is then related to the length of the gun, $\ell = \frac{1}{2} a t_f^2$.  The time $t_o(x)$ at which the armature passes coordinate $x$ is then  $t_o(x)=(2 x/a)^{1/2}$, see also Eq.~(\ref{constantCurrent}).  Using these approximations in Eq.~(\ref{tempRiseDistributionConstantCurrent_1}) gives the temperature rise at position $(x,y)$
\begin{equation}
\begin{array}{l}
 T(x,y,t_f ) - T_0  = \frac{{\mu \,I_0 ^2 }}{{4w^2 \rho \,C}}\;\; \times \; \\ 
 \left\{ {\begin{array}{*{20}c}
   {Log\left[ {\frac{{8\, \ell}}{{\mu \,\sigma \,v\,y^2 }}\left( {1 - \left( {\frac{x}{\ell}} \right)^{1/2} } \right)} \right],} & {y\, \le \,\,\tilde y(x)}  \\
   {0,} & {y > \,\tilde y(x)}  \\
\end{array}} \right. \\ 
 \end{array}
\label{temprise3}
\end{equation}
where 
\begin{equation}
\tilde y(x) = \left( {\frac{{8\, \ell }}{{\mu \,\sigma \,v\,}}} \right)^{1/2} \,\left( {1 - \left( {\frac{x}{ \ell }} \right)^{1/2} } \right)^{1/2} 
\label{ytildaDef}
\end{equation}
and $x$ is assumed to be in the interval $0<x\le \ell $.  Note that Eq.~(\ref{temprise3}) predicts that the temperature on the surface of the rail, at $y=0$, is infinite. This feature of the solution is well-known and is not a problem~\cite{Knoepfel1970}. The singularity with respect to $y$ is integrable, and therefore the energy deposited in a thin layer near the surface is finite.  We use Eq.~(\ref{temprise3}) to locate the points $(x,y)$ of the surface that reaches the melting point.     

Using the values for the gun parameters and material parameters in Tables~I and II, we plot the distribution of the temperature rise in the rail, $T(x,y,t_f ) - T_0$, as a function of the $y-$coordinate, for several points along the rail given by coordinate $x$.  For comparison, we also plot the melting point of Cu and Ta, which is 1083$^\circ C$ and 2996$^\circ C$, respectively.   The $y-$coordinate where the curve intersects the melting point of each metal is the depth to which melting of the rails occurs, if the EMG gun shot occurred with rails at initial temperature $T_0$.   As mentioned previously, we have not taken into account the heat of melting, so these curves are not correct for temperatures above the melting point of the rails.

\begin{figure}[tbp] 
  \centering
  \includegraphics[width=2.5in]{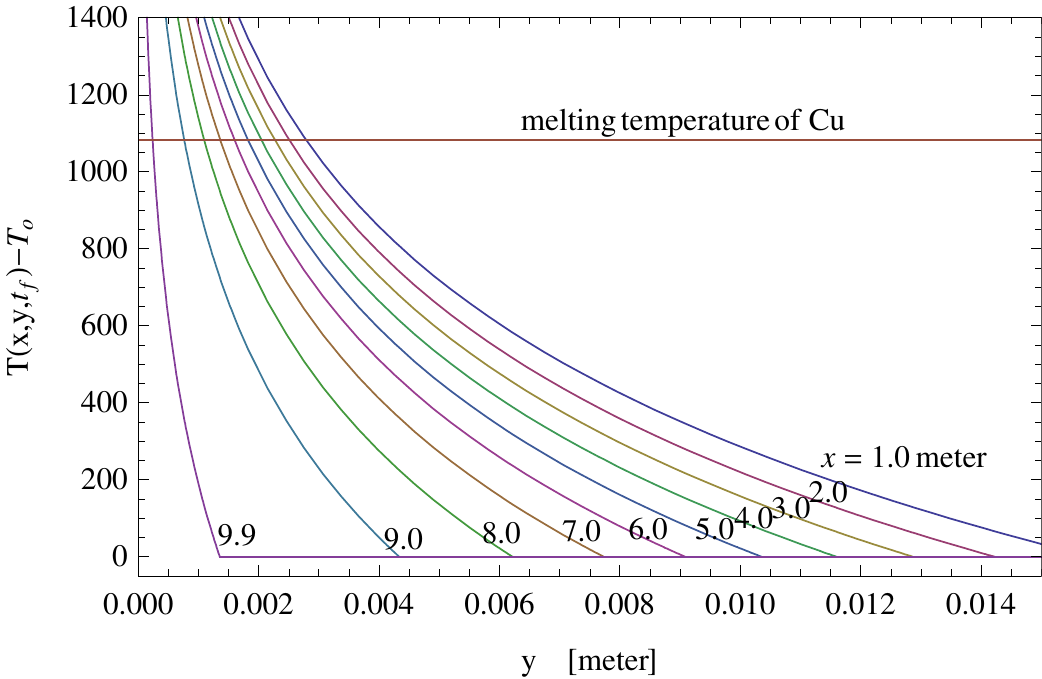}
  \caption{Plot of Eq.~(\ref{temprise3}), giving  the temperature rise, assuming rails made from copper, plotted vs. position $x$ along the rail. Parameters used are given in Tables I and II.  The gun velocity is taken as 3000 m/s and the inital temperature $T_o=0^{o}$C.  Constant current is assumed.}
  \label{fig:EMRG_Figure6_Cu_temp_rise}
\end{figure}
Figure~6 and 7 show a plots of the thickness of the melted layer assuming the rails are made of copper and tantalum, respectively, for a constant total gun current, which corresponds to a unifomly accelerating armature.
\begin{figure}[tbp] 
  \centering
  \includegraphics[width=2.5in]{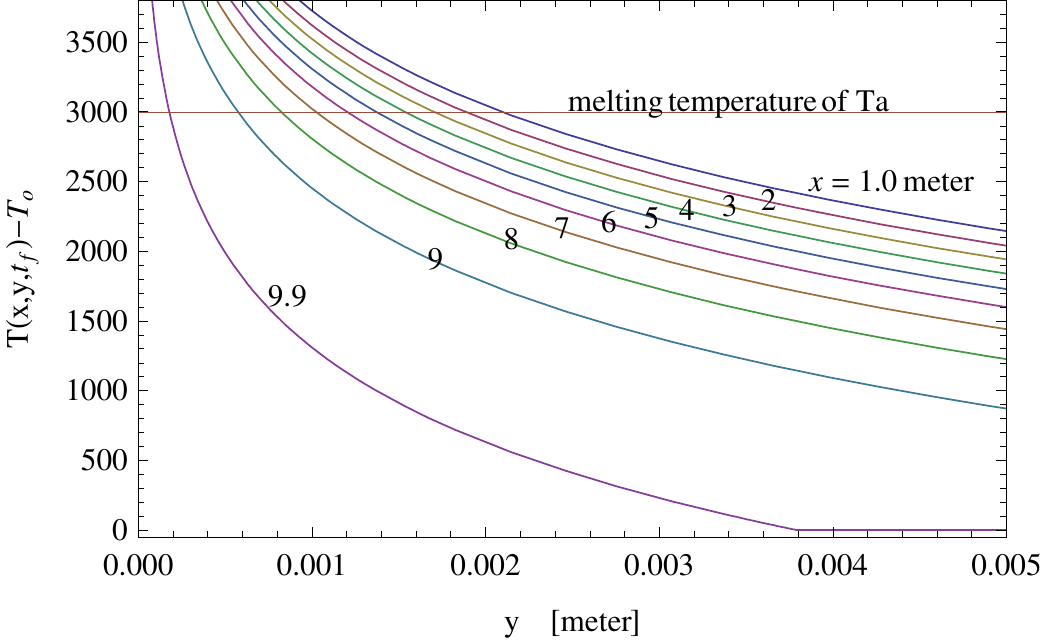}
  \caption{Plot of Eq.~(\ref{temprise3}), giving  the temperature rise, assuming rails made from tantalum, plotted vs. position $x$ along the rail. Parameters used are given in Tables I and II.  The gun velocity is taken as 3000 m/s and the inital temperature $T_o=0^{o}$C.  Constant current is assumed.}
  \label{fig:EMRG_Figure7_Ta_temp_rise}
\end{figure}

Our model predicts that the thickness of the melted layer is largest at $x=0$, at the breech of the gun rail.  This is reasonable because the current at the breech of the gun flows for the longest time, causing maximum Joule heating at $x=0$.

The intersection of the curves in Figure~6 and 7 with the melting point of metal  (horizontal line) gives the thickness of the melted layer. The surface inside the rails that reaches the melting temperature, $T_{melt}$, is given by points $(x,y)$ that satisfy 
\begin{equation}
T(x,y,t_f ) - T_0 = T_{melt} -T_0
\label{melting_surface}
\end{equation} 
Alternatively, we can say that at position $x$ the thickness of the rail that reaches the melting point, $y_{melt}(x)$,  is given implicitly by $T(x,y_{melt},t_f ) - T_0 = T_{melt} -T_0$.  For any element of volume, the temperature rise is due to the length of time that the current was flowing through that element.  For an element at $(x,y)$, current starts to flow at time $\tau(x,y)$. We assume that for all elements the current stops flowing at time $t_f$, when the armature leaves the rails. 

\begin{figure}[tbp] 
  \centering
  \includegraphics[width=2.5in]{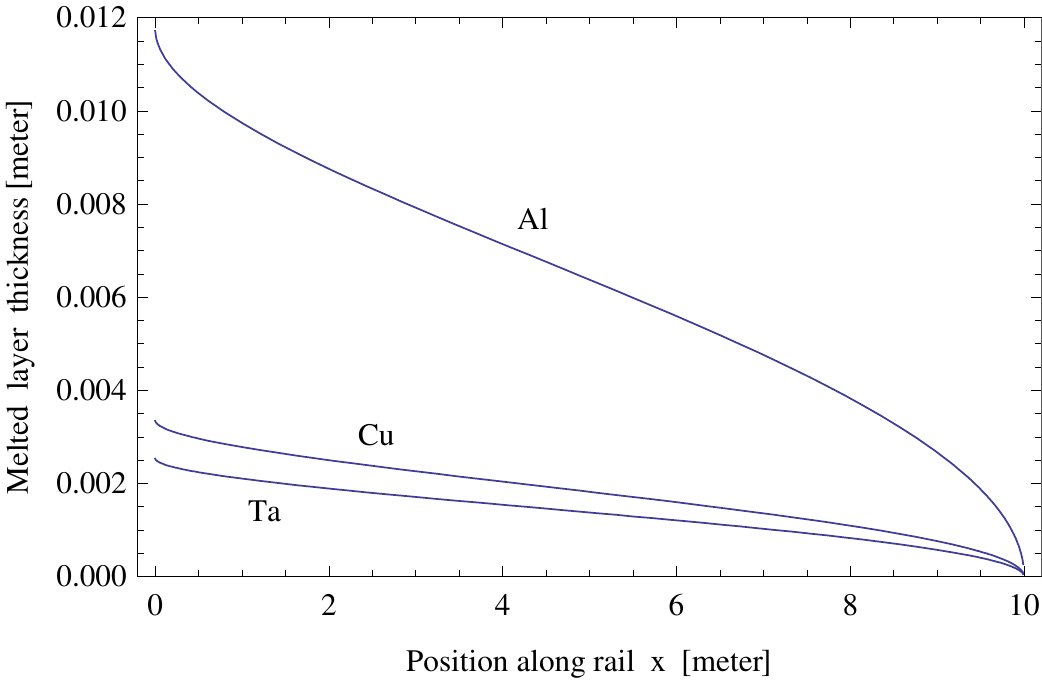}
  \caption{Plot of Eq.~(\ref{melt_thickness}), giving the melt layer thickness vs. x along the rails for rails made from copper, tantalum and aluminum. The initial temperature of rails is assumed to be $T_o=0^{o}$C.  Constant current is assumed and the gun velocity is taken to be 3000 m/s.}
  \label{fig:EMRG_Figure8_Melt_Thickness_Cu_Al_Ta}
\end{figure}

From Eq.~(\ref{melting_surface}) we solve for the thickness of the melted layer, $y_{melt}(x)$, as a function of $x$ position along the rail
\begin{equation}
y_{melt}(x)  = \left( {\frac{{8\, \ell}}{{\mu \,\sigma \,v}}} \right)^{1/2} \,\;e^{ - \frac{{2\,{\kern 1pt}  w^2 \rho {\kern 1pt} {\kern 1pt} \, L' \, \ell \, C {\kern 1pt}  \left( {T_{{\rm{melt}}}  - T_o } \right)}}{{\mu \,{\kern 1pt} m \, v^2 }}} \,\,\left[ {1 - \left( {\frac{x}{\ell}} \right)^{1/2} } \right]^{1/2} 
\label{melt_thickness}
\end{equation}
Figure~\ref{fig:EMRG_Figure8_Melt_Thickness_Cu_Al_Ta} shows a plot of the thickness of the melted layer assuming the rails are made of aluminum, copper, and tantalum.  We used numerical values given in Tables I and II.

\begin{figure}[tbp] 
  \centering
  \includegraphics[width=2.5in]{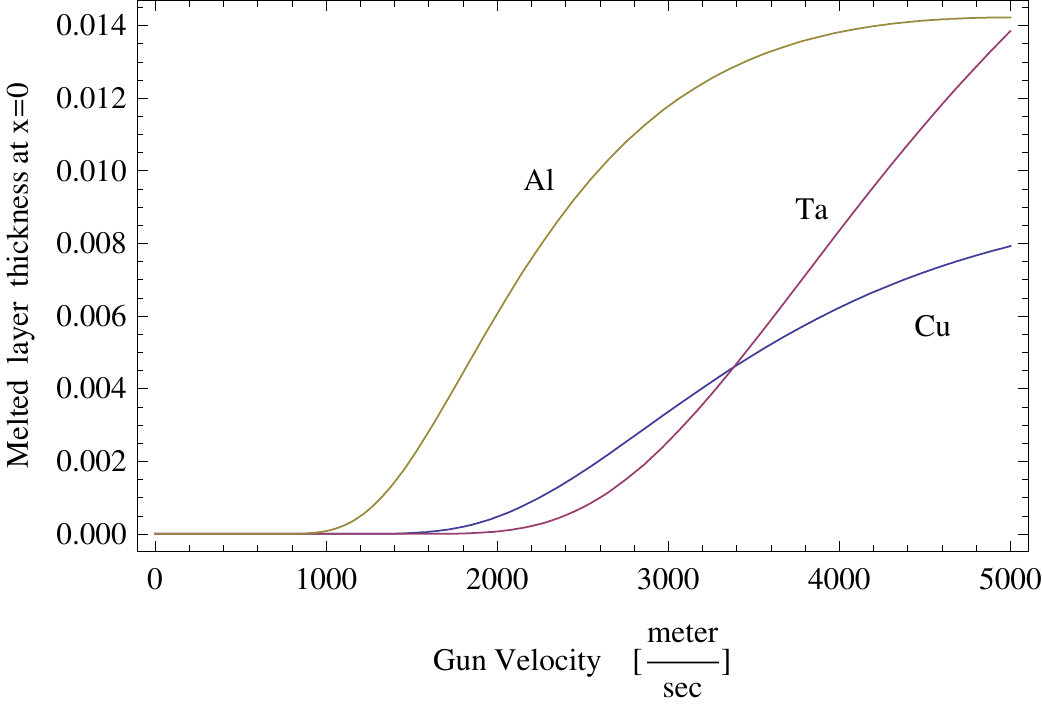}
  \caption{Plot of Eq.(\ref{melt_thickness}) giving the melted layer thickness for Cu, Ta, and Al rails at the gun breech (at x=0) vs. gun velocity, $v$. The initial temperature of rails is assumed to be $T_o=0^{o}$C. Constant current is assumed.  This plot also gives an estimate of the melted layer thickness in the armature, see discussion in Section VI.}
  \label{fig:EMRG_Figure9_Melt_Thickness_Cu_Al_Ta_vs_Vel}
\end{figure}

The thickness of the melted layer has a strong dependence on gun velocity, which is the velocity of the armature at $x=\ell$.  Figure~\ref{fig:EMRG_Figure9_Melt_Thickness_Cu_Al_Ta_vs_Vel} shows plots of the melted layer thickness at $x=0$ for Cu, Ta, and Al rails  vs. gun velocity.
The plots in Figure~\ref{fig:EMRG_Figure9_Melt_Thickness_Cu_Al_Ta_vs_Vel} show that for gun velocities above 1500 m/s, using materials such a copper, aluminum,  and tantalum, the thickness of the melted layer in the gun rails increases rapidly with gun velocity.   
\begin{table}
\caption{\label{gun_parameters}Gun parameters.}
\begin{tabular}{lll}
quantity & symbol &  value \\
\hline
width of rails & $w$ & 0.10 m \\
length of rails (gun length) &  $\ell$  &  10.0 m \\
mass of armature  & $m$ &  20 kg  \\
derivative of self inductance of rail  &  $L'$   &  0.46$\times$10$^{-6}$ H/m  \\
magnetic permeability   &  $\mu$    &  $4 \pi \times 10^{-7}$  H/m 
\end{tabular}
\end{table}
\begin{table}
\tiny
\caption{\label{material_parameters}Material parameters for the rails.}
\begin{tabular}{ccccc}
 symbol   &   $T_{melt}$    &  $\sigma$    &   $\rho$   &  $C$  \\
\hline
       &   melting temp. ${}^oC$ &  conductivity (Ohm m)$^{-1}$ & density kg/m$^{-3}$ &  specific heat J/kg ${}^o C$  \\
\hline
aluminum   &     660   & 3.82 $\times$ 10$^7$    & 2720.  & 950. \\
copper     &    1084   & 5.8  $\times$ 10$^7$    & 8960.  & 440.  \\
tantalum   &    2996   & 7.40$\times$10$^6$     & 16600. & 150.62 
\end{tabular}
\end{table}
\section{Current Distribution due to Magnetic Field Diffusion} 

In this section we obtain the current density in the rail, $J(x,y,t)$ (to be used in Eq.~(\ref{heating})), by considering the magnetic field diffusion into the rails. We treat the diffusion of the magnetic field into the rails as diffusion into a plane surface, with a time dependent boundary condition on the field, given by the armature passing a surface element of the rail.   Consider an arbitrarily thin armature at position $x_o$ at some time $t$.  For the geometry in Figure 1, the magnetic field $H_z$ is in the $z$-direction.   For $x>x_o$, which is outside the rail-armature circuit, the magnetic field $H_z=0$.  For $0<x<x_o$, which is inside the rail-armature circuit,  the magnetic field has some constant value $H_0$.   As the (arbitrarily thin) armature passes the point $x$ on the surface of the rail, the surface magnetic field changes from $H=0$ to some finite value $H_0$, and the field starts to diffuse into the rail. Essentially, as the armature passes the point $x$  the boundary condition on the field $H_z$ on the rail surface changes from $H_z=0$ to 
$H_z=H_0$.    Considering the rail to be a plane surface, the magnetic field diffuses into the rail according to 
the equation~\cite{LL_continuous_media,Knoepfel1970} 
\begin{equation}
\frac{{\partial ^2 H_z }}{{\partial y^2 }} - \frac{1}{\kappa }\frac{{\partial H_z }}{{\partial t}} = 0
\label{Hdiffusion}
\end{equation}
For a plane surface with magnetic field boundary condition $H_z=0$ for $t<0$ and $H_z=H_0$ for $t>0$, the solution for $y>0$ is 
\begin{equation}
H_z  = H_0 \,{\rm{erfc}}\left( {\frac{y}{{2\sqrt {\kappa {\kern 1pt} t} }}} \right)
\label{Hsolution}
\end{equation}
where $\kappa = 1/(\mu \sigma)$ is the magnetic diffusion length, ${\rm{erfc}}\left(\xi\right)$ is the complimentary error function, ${\rm{erfc}}\left(\xi\right)= 1- {\rm{erf}}\left(\xi\right)$ and ${\rm{erf}}\left(\xi\right)$ is the error function given by $$ erf (z)=\frac{2}{\sqrt{\pi }}\int _0^ze^{-t^2}dt $$ 
The current density associated with this magnetic field can be obtained from Eq.~(\ref{MaxwellCurrentDensity}), leading to
\begin{equation}
{\bf{J}} = {\bar J}(y,t) \,\;{\bf{e}}_x  = \frac{{H_0 }}{{\sqrt {\pi {\kern 1pt} \kappa {\kern 1pt} t} }}\,\exp \left( { - \frac{{y^2 }}{{4\,\kappa \,{\kern 1pt} t}}} \right)\,\,{\bf{e}}_x 
\label{Jcurrent}
\end{equation}
where ${\bf{e}}_x$ is the unit vector in the $x$-direction, and the auxiliary function, ${\bar J}(y,t)$, is defined by Eq.~(\ref{Jcurrent}). The value of the magnetic field is related to the total current $I_o$ in the rails by 
\begin{equation}
I_o  = \int {\int {dy\,dz\,} } {\bar J}(y,t) = w H_0
\label{Total}
\end{equation}
leading to $H_0 = I_o/w$ where $w$ is the width of the rail in the $z$-direction. 

We use the plane surface solution to approximate the magnetic field diffusion and resulting current distribution in the rail.  As the armature sweeps past a surface element in the rail at position $x$, the field starts diffusing into the rail at time $t_o(x)$, where   $t_o(x)$ gives the time the armature passes point $x$.  We take the current density to have the form
\begin{equation}
{\bf{J}}(x,y,z,t) = \bar J(y,t - t_o (x))\,{\bf{e}}_x 
\label{JcurrentDEF}
\end{equation}
when the coordinates $(x,y,z)$, and time $t$ satisfy the conditions
\begin{equation}
0 < x < x_o (t)\, \,\: , \,\,y > 0\,\: ,  \,\,\,\:|z| < \frac{w}{2} \,\, \, ,  \,\,\; 0 < t < t_f  
\label{coordinateConditions}
\end{equation}
and otherwise we take ${\bf{J}}(x,y,z,t) =0$.
Here, $x_o (t)$ is the $x$-position of the armature at time $t$, and ${\bar J}(y,t)$ is given by Eq.~(\ref{Jcurrent}). The quantity ${\bf J}(x,y,z,t))$ is an approximation to the time-dependent current density in the rails during the EMG shot.   Note that 
${\bf J}(x,y,t)$ does not depend on $z$ because we assume a large (essentially infinite) extent in the $z$-direction. 

The temperature rise is obtained from  Eq.~(\ref{heating}) and we neglect the effect of re-distribution of heat during the time of the shot, since it takes a long time on the time scale of a shot (which is $t_f$),  The temperature rise  immediately after the shot is
\begin{equation}
T(x,y,t_f ) - T_0  = \frac{{\mu \,I_o ^2 }}{{\pi {\kern 1pt} {\kern 1pt} w^2 {\kern 1pt} \rho {\kern 1pt} {\kern 1pt} C}} \,\,\, \Gamma \left( {0,\frac{{\mu {\kern 1pt} \sigma }}{2}\frac{{y^2 }}{{t_f  - t_o (x)}}} \right)
\label{tempRiseDistribution}
\end{equation}
where $\Gamma(a,z)$ is the incomplete Gamma function, given by 
$$\Gamma (a,z)=\int _z^{\infty }t^{a-1}e^{-t}dt$$  
Equation~(\ref{tempRiseDistribution}) is our basic result for the temperature distribution in the rails, and is expressed in terms of the current in the rails $I_o$.  It is useful to express the temperature rise in terms of the gun velocity $v$.

In order to get simple results, we assume that the acceleration is constant, using the relations in Eq.~(\ref{constantCurrent})
leads to the temperature rise
\begin{equation}
T(x,y,t_f ) - T_0  = \frac{{\mu {\kern 1pt} {\kern 1pt} m{\kern 1pt} {\kern 1pt} v^2 }}{{\pi \,w^2 \,\rho \,C\,L'\, \ell}}\;\Gamma \left( {0,\frac{{\mu \,\sigma \,v}}{{4\,\sqrt \ell }}\,\frac{{y^2 }}{{\sqrt \ell  - \sqrt x }}} \right)
\label{tempRiseDistributionConstantCurrent}
\end{equation}
Figure~12 shows a plot of the temperature rise given by Eq.~(\ref{tempRiseDistributionConstantCurrent}), for the case of copper rails, using the parameters in Tables~I and II.
\begin{figure}[tbp] 
  \centering
  \includegraphics[width=2.5in]{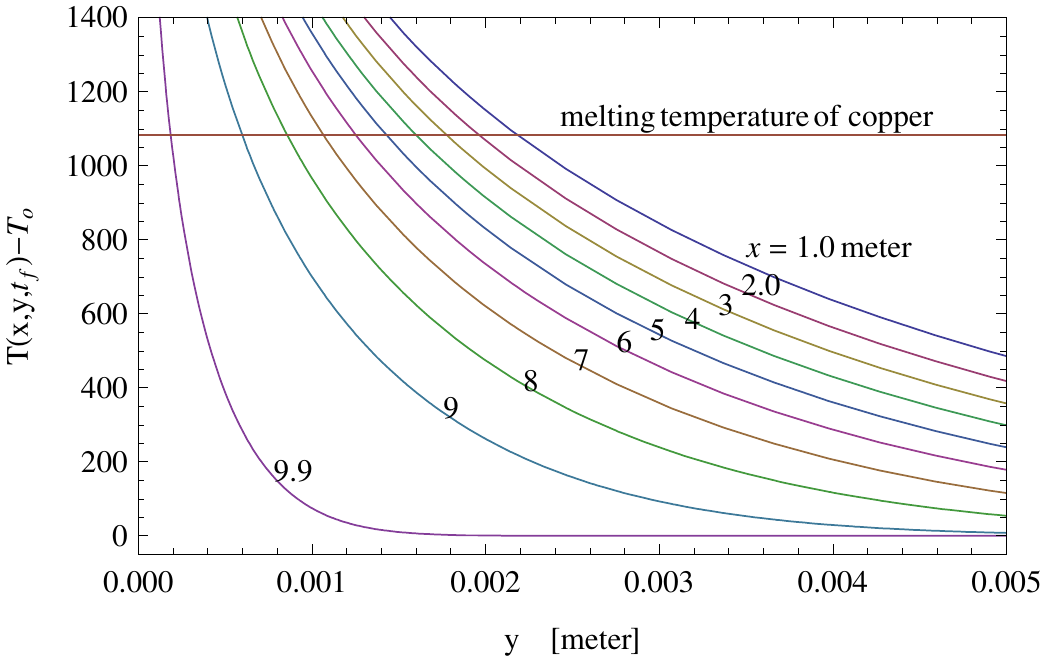}
  \caption{The temperature rise is plotted vs. $y$ coordinate into the rail for various positions $x$ along
the copper rail. The time dependence of the current is determined by the time dependence of magnetic field diffusion into the rails.    Initial temperature of the rails was assumed to
be $T_0$ =0 C. Parameters used are given in Tables~\ref{gun_parameters}  and \ref{material_parameters}.}
  \label{Temp_rise_field_diffussion}
\end{figure}
\begin{figure}[tbp] 
  \centering
  \includegraphics[width=2.5in]{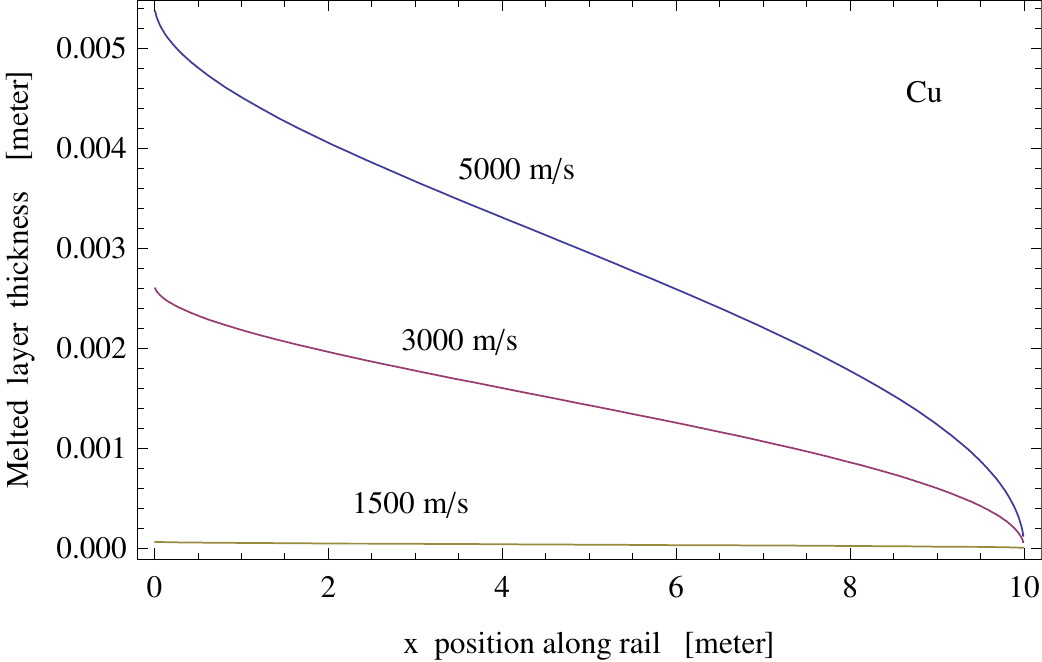}
  \caption{Plot of the thickness of the melted layer vs. position x along the rail, assuming rails made from copper.  Curves for three different velocities are shown. Initial temperature of the rails was assumed to be $T_0$ =0 C. Parameters are those in Tables~\ref{gun_parameters}  and \ref{material_parameters}.}
  \label{fig:EMRG_melt_thickness_Cu_Figure11_field_diffussion}
\end{figure}
\begin{figure}[tbp] 
  \centering
  \includegraphics[width=2.5in]{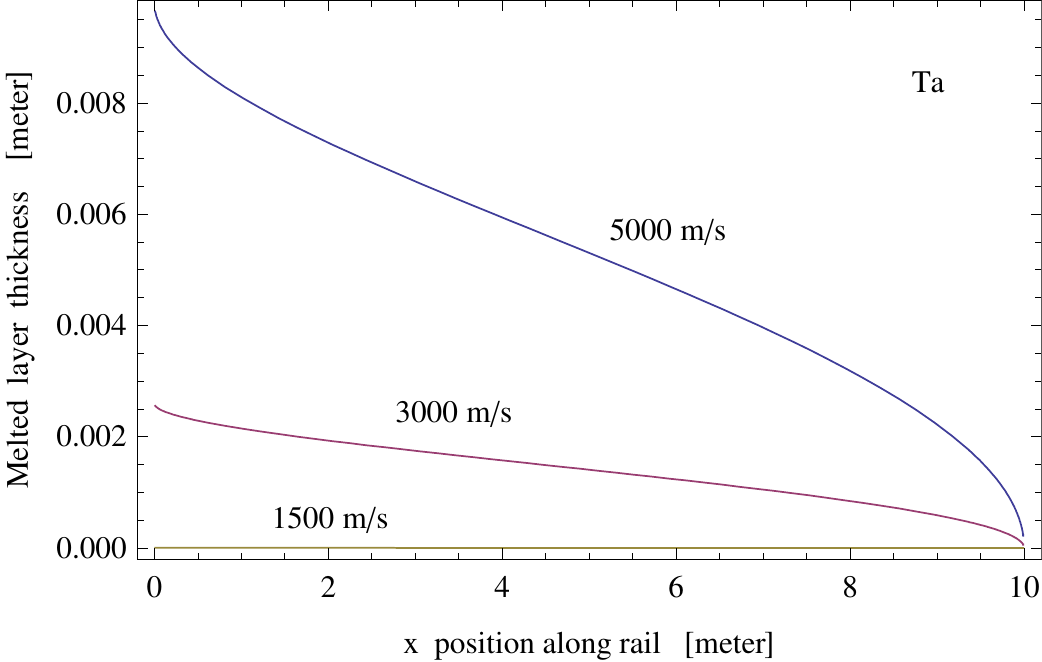}
  \caption{Plot of the thickness of the melted layer vs. position x along the rail, assuming rails made from tantalum.  Curves for three different velocities are shown. Initial temperature of the rails was assumed to be $T_0$ =0 C. Parameters are those in Tables~\ref{gun_parameters}  and \ref{material_parameters}.}
  \label{fig:EMRG_melt_thickness_Ta_Figure12_field_diffussion}
\end{figure}%

In Eq.~(\ref{tempRiseDistributionConstantCurrent}),  the argument of the $\Gamma(\xi)$ function is large for all values of $x$ and $y$, except for $y$ near $y=0$. For the values in Tables I and II, the argument of the $\Gamma$ function is
\begin{equation}
\xi = \frac{v y^2 \mu  \sigma }{4 \left( \ell -\sqrt{\ell} \sqrt{x}\right)}=\frac{54663.71 \, \, y^2}{10.-3.1622 \sqrt{x}}
\label{xi}
\end{equation}
Therefore, the $\Gamma(\xi)$ function for $\xi >>1$ can be approximated by its asymptotic expansion
\begin{equation}
\Gamma(0, \xi ) = e^{-\xi } \left(\frac{1}{\xi }-\frac{1}{\left(\xi \right)^2}+O\left[\frac{1}{\xi }\right]^3\right)
\label{Gamma_asymptotic}
\end{equation}%
Near y=0, the behavior of the $\Gamma (\xi)$ function for $\xi <<1$ is logarithmic
\begin{equation}
\Gamma(0, \xi ) =  (-\gamma-  Log [\xi ])+\xi -\frac{\xi ^2}{4}+O[\xi ]^3`
\label{xi_small_arg}
\end{equation}
where  $ \gamma$ is Euler's constant, $ \gamma \approx 0.5772$.

As before, see Eq.~(\ref{melting_surface}), we use Eq.(\ref{tempRiseDistributionConstantCurrent}) to solve for the thickness of the melted layer in the $y$-direction by setting the temperature rise equal to the melting temperature of the metal.  Figures~11 and 12 show plots of the thickness of the melted layers verses position $x$ along the rail, for several velocities, for rails made of copper and tantalum, respectively. From these figures, we see that even though Ta has a much higher melting temperature than Cu, 2996${}^o$C  compared to 1084${}^o$C, respectively, the electrical conductivity of Ta is lower, which leads to a comparable thickness for the melted layer for Ta and Cu rails.

\section{Armature Heating}
We can estimate the heating of the armature from the calculations that we have already done. We now imagine that the armature has a finite thickness.  Furthermore, depending on the EMG design, the armature material can be different than that of the rails, and consequently, the electrical conductivity $\sigma$ may be different.  Unlike the rails, the armature conducts current for the whole duration of the shot from $t=0$ to $t=t_f$.  In this sense, the whole length of the armature has current flowing in it like the element of rail at coordinate $x=0$.   The thickness of the melted layer of the armature can be found from Eq.(\ref{melt_thickness}) by setting $x=0$,
\begin{equation}
y_{melt}(0)  = \left( {\frac{{8\, \ell}}{{\mu \,\sigma \,v}}} \right)^{1/2} \,\;e^{ - \frac{{2\,{\kern 1pt}  w^2 \rho {\kern 1pt} {\kern 1pt} \, L' \, \ell \, C {\kern 1pt}  \left( {T_{{\rm{melt}}}  - T_0 } \right)}}{{\mu \,{\kern 1pt} m \, v^2 }}}  
\label{armature_melt_thickness}
\end{equation}  
Figure~9 shows a plot of the melted layer thickness of the armature verses gun velocity, as given by Eq.~(\ref{armature_melt_thickness}).

\section{Efficiency and Melt Layer Thickness}
During an EMG shot, the energy that is supplied to the gun appears as projectile kinetic energy, Joule heating of the rails and armature, armature and rail heating caused by frictional forces between rails and armature, vibration of rails and armature, sound and light produced in surrounding air, and electromagnetic radiation due to a time-dependent magnetic field. We neglect all these effects except for the Joule heating of the rails and kinetic energy of the projectile and armature. 

From the current density in Eq.~(\ref{JcurrentDEF}), we calculate the energy in {\it one} rail, $Q$, that is dissipated over one shot of the EMG by integrating the Joule heating over the time $t_f$ during which the armature is in contact with the rails,
\begin{equation}
Q = 2 \int {d^3 x} \int\limits_0^{t_f } {dt} \,\frac{{J^2 }}{\sigma }
\label{Joule_Heat_def}
\end{equation} 
where the spatial integration is over all space inside one of the rails.  The factor of two is due to the fact that there are two rails, and by symmetry the energy dissipated is twice that of one rail.  Assuming constant current, for which the time $t_f$ is defined in Eq.~(\ref{constantCurrent}),  assuming the conductivity  $\sigma$ is a constant, and using the expression given by Eq.~(\ref{JcurrentDEF}) for the current density derived by considering magnetic field diffusion into rails each of length $\ell$, we find
\begin{equation}
\begin{array}{c}
 Q = \frac{{32}}{{15}}\left( {\frac{\mu }{{\pi \sigma }}} \right)^{1/2} \,\frac{{\ell^{5/4} \,\,m^{1/4} }}{{L^{\prime 1/4} \,\,w}}\,\,I_o^{3/2} \, \\ 
  = \,\frac{{32}}{{15}}\left( {\frac{\mu }{{\pi \sigma }}} \right)^{1/2} \,\frac{m}{{w\,L^\prime  }}\, \ell^{1/2} \,\,v^{3/2}  \\ 
 \end{array}
\label{Joule_Heat}
\end{equation} 
where we have assumed a constant current (armature acceleration) and used the value of $t_f$ in Eq.~(\ref{constantCurrent}).  Note that $Q$ is the total amount of energy dissipated in both rails.   The two forms for $Q$ are obtained assuming the relation between EMG velocity and current given in Eq.~(\ref{constantCurrent}), which assumes a constant current.   Equation~(\ref{Joule_Heat}) gives the amount of energy $Q$ dissipated in the rails due to Joule heating over the time $t_f$ of one shot.  Equation~(\ref{Joule_Heat}) contains some interesting physics because the total energy dissipated as Joule heat is not proportional to $I_o^2$ as is the case for a resistor carrying a constant current $I_o$ for a time $t_f$.  If the rails acted as a simple resistor with resistance $R$, then the amount of energy dissipated due to Joule heating during the shot would be $I_o^2 \, R \, t_f$.  If $t_f$ was a constant independent of current $I_o$, then the amount of heat dissipated in the resistor would be proportional to $I_o^2$.  However, in an EMG, the time $t_f$ is inversely proportional to current $I_o$, as given by Eq.~(\ref{constantCurrent}), because the armature spends less time in contact with the rails at higher current (due to higher average armature velocity at higher current). So if the rails acted like a resistor with no skin effect, and the time $t_f$  varied inversely with current $I_o$, then the amount of heat dissipated in such a  resistor would be proportional to $I_o$.   However, in the EMG model that we are considering, we have included the skin effect.  With increasing current $I_o$, the time $t_f$ is shorter and there is a stronger skin effect at larger $I_o$ leading to an increased resistance $R$, resulting in an increased Joule heat dissipation during the shot time $t_f$, since all current has to flow through a thinner channel in the rails. The net effect is that the skin effect causes an increased Joule heat over the shot, so that for the EMG with skin effect we have $Q \sim I_o^{3/2}$, as given in Eq.~(\ref{Joule_Heat}).  

The ratio of armature kinetic energy divided by the Joule heat is a sort of ``thermal $q$'' and is given by   
\begin{equation}
\begin{array}{c}
 q   =    \frac{{\frac{1}{2}mv^2 }}{Q} \\ 
      =  \frac{{15}}{{32}}\left( {\frac{{\pi \sigma }}{\mu }} \right)^{1/2} \frac{{{\rm{w}}\,L'^{{\kern 1pt} {\kern 1pt} 5/4} }}{{\left( {m{\kern 1pt} \ell } \right)^{1/4} }}\,I_o ^{1/2}  \\ 
      =  \frac{{15}}{{32}}\left( {\frac{{\pi \sigma }}{\mu }} \right)^{1/2} \,{\rm{w}}\,L'\,\left( {\frac{v}{\ell}} \right)^{1/2}  \\ 
 \end{array}
\label{Ratio_KE_to_Joule_Heat}
\end{equation} 
The quantity $q$ shows that, in order to achieve a certain velocity $v$, a larger fraction of input energy  goes into the armature kinetic energy when using shorter rails (smaller $\ell$). Figure~15 shows the dependence of $q$ on velocity $v$ for parameters given in Table I and II for gun rails made of copper.
\begin{figure}[tbp] 
  \centering
  \includegraphics[width=2.5in]{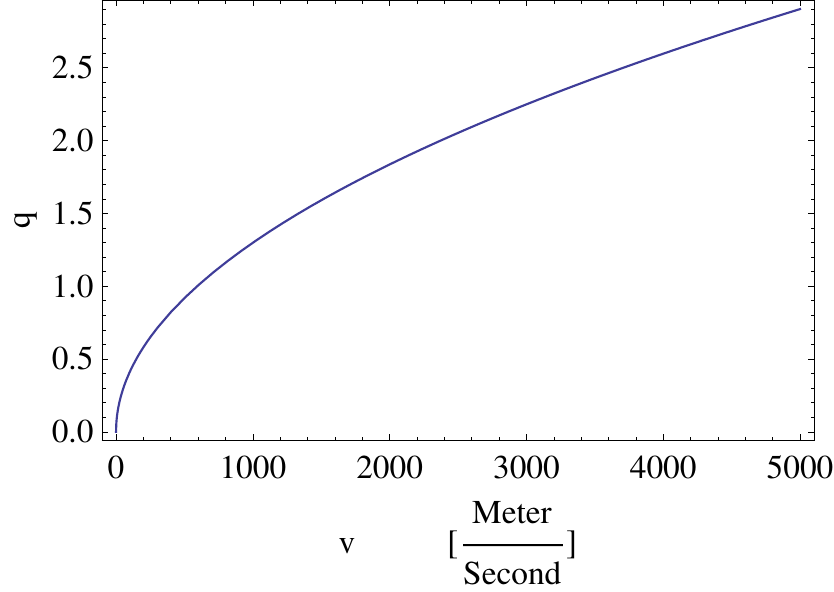}
  \caption{The plot shows the thermal $q$ defined in Eq.~(\ref{Ratio_KE_to_Joule_Heat}) as a function of velocity, using the parameters in Table I, assuming copper rails with conductivity in Table II. }
  \label{fig:EMRG_thermal_q_Figure13}
\end{figure}
However, the quantity $q$ does not contain the whole story.  We next consider the EMG efficiency.

Within our model, we define the efficiency of an EMG, $\eta$, as the armature kinetic energy, divided by the total energy
\begin{equation}
\eta  = \frac{{\frac{1}{2}mv^2 }}{{\frac{1}{2}mv^2  + Q + E_{mag}(\ell) }}
\label{efficiency_definition}
\end{equation}
We consider the total energy as a sum of three terms: the kinetic energy of the armature $\frac{1}{2}m v^2$, the energy dissipated by Joule heating $Q$ in Eq.~(\ref{Joule_Heat}), and the energy stored in the magnetic field $E_{mag}(\ell)$ when the armature is at the end of the rails at $x=\ell$. In general, the magnetic energy $E_{mag}(x)$ is the energy stored in the magnetic field in the rails and surrounding space when the armature is at position $x$:
\begin{equation}
E_{mag} (x) = \frac{1}{2}\, L(x) \, I_o^2 
\end{equation}   
 As previously defined in Eq.~(\ref{variableInductance}), we write $ L(x)$ in terms of $L_0$ and $L^\prime$, where $L_0$ is the self inductance of the EMG that does not depend on armature position (and corresponds to magnetic energy $L_0 \, I_o^2 \, /2$ stored in the power supply, surrounding space, and in electrical leads to the rails)  and $L^\prime$, which is the derivative of the self inductance with respect to armature position $x$.  When the armature reaches the end of the rails,  the circuit is broken and the magnetic energy $E_{mag}(\ell)$ is dissipated in the form of an electrical arc, sound, light, and mechanical vibration. Assuming constant current and acceleration, when the armature is at the end of the rails at position $x=\ell$, the stored magnetic energy is 
\begin{equation}
E_{mag} (\ell) = \frac{1}{2}\,mv^2 \left( {1 + \frac{{L_0 }}{{L'\; \ell}}} \right)
\label{magnetic_energy_at_L}
\end{equation}
where we have used Eqs.~(4).  From Eqs.~(30),~(\ref{efficiency_definition}) and (\ref{magnetic_energy_at_L}), the efficiency of the EMG is then 
\begin{equation}
\eta  = \frac{1}{{2 + \frac{{L_0 }}{{L'\; \ell}} + \frac{{64}}{{15\,w{\kern 1pt} L'}}\left( \frac{\mu \, \ell }{\pi \sigma \,v} \right)^{1/2} \,\;\,}}
\label{efficiency}
\end{equation} 
Equation~(\ref{efficiency}) shows how the efficiency depends on a number of variables. The efficiency of an EMG increases with  increasing electrical conductivity $\sigma$, velocity $v$, and rail width $w$. The efficiency depends in a complicated way on the rail length $\ell$.  In the limit of high velocity $v$, the efficiency of the EMG has the limiting value
\begin{equation}
\mathop {\lim }\limits_{v \to \infty } \;\eta  = \frac{1}{{2 + \frac{{L_0 }}{{L'\;\ell}}\,}}
\label{efficiency_limit}
\end{equation}
The high-velocity limit of EMG efficiency depends on the ratio of the stationary self inductance, $L_0$, to the dynamic part of the self inductance of the rails, ${L'\; \ell}$.  We call ${L'\; \ell }$ the dynamic part of the self inductance because it depends on the length of the rails.   For a given gun design, the quantities $L_0$ and ${L'}$ are constants, but in principle the length of the rails, $\ell$, can be increased to make the term $\frac{L_0}{L' \; \ell } <<1$, thereby increasing the high-velocity limit of the efficiency.    Longer rails (larger $\ell$) lead  to a higher EMG efficiency at high velocity.  However, for all velocities the EMG efficiency for this model is always less than 1/2.   

The denominator of the efficiency in Eq.~(\ref{efficiency}) has three terms. For the parameters in Table I and II, the third (last) term in the denominator is  $24.35 / \sqrt(v)$, where $v$ is in units of m/s.  Therefore, it is clear that the EMG is more efficient at higher velocities.  The efficiency will increase with  velocity significantly when the stationary part of the self inductance, $L_0$, is smaller than the dynamic part of the self inductance, $L' \; \ell $,  so that the term $\frac{{L_0 }}{{L'\; \ell}}<<1$, which may rarely be true in real designs unless the length of the rails $\ell$  can be made sufficiently large.  

Stated in another way, a longer EMG is more efficient at higher velocities,  however, the thickness of the melted layer also increases at higher velocities.  Therefore, there is a tradeoff between efficiency and melting of the rails.  Increasing the length of the rails may prevent melting, however,  this may lead to rail length that is not practical for applications.

In our simple model of an EMG, and our definition of efficiency, we have neglected many effects such as friction between armature and rails, plasma contacts, and many other details.  These effects would make the denominator of Eq.~(\ref{efficiency}) larger, and hence would reduce the efficiency.

\section{Conclusion}  
We have constructed a model of the EMG based on the electrodynamics of the launch rails and armature. Our model takes into account the skin effect in the rails and armature, which is significant due to the short-duration and  extremely high current densities in high-performance EMGs. We used two approaches.  The first approach, in Section III we modeled the skin effect by a constant-in-time current flowing through a channel equal to the local skin depth along the rail, $\delta$, given by Eq.~(7) at each point along the rail.   We find that a finite thickness layer of the rails, given by Eq.~(\ref{melt_thickness}), reaches the melting point of the metal.   Although aluminum is lighter by a large factor, we have assumed the rails are made from copper because of its significantly higher melting point (1083 C for copper verses 660 C for aluminum).  Figures~6 and 7 show plots of the temperature rise (immediately after a shot) versus depth into the rails, for different positions spaced one meter apart along the rails, assuming 10 meter long rails, for copper and tantalum rails, respectively. For all positions along the rails, immediately after the shot and before thermal diffusion takes place, the temperature rise is the highest at the rail surface and decreases into the interior of the rails.  For constant current, near the breech end of the gun the skin depth exceeds 15 mm, and a layer approximately 2~mm or 3~mm thick melts, for tantalum or copper  rails, respectively, see Figure~8.  The model predicts that, immediately after a shot and before thermal energy is redistributed, the temperature is largest near the rail surface and a thin surface melt layer exists for any rail gun, see Eq.~(\ref{melt_thickness}).   

In the second approach, in Section IV, we obtained an approximate current density by computing the time dependence of the diffusion of the magnetic field into the rail surface during the EMG shot. From the magnetic field ${\bf H}$, we obtain the current density from Eq.~(\ref{MaxwellCurrentDensity}), which is used to compute the temperature rise immediately after a EMG shot, see Figures~6, 7 and 10.   Figures~11 and 12 show the melted layer thickness we can expect for copper and tantalum rails, for different gun velocities (due to different gun  currents, which are related to velocities by Eq.~(4)).  We expect this approach (Section IV) to be our most accurate estimation of the temperature rise in the rails and resultant melting.  In both approaches, we find the thickness of the melted layer of the rail increases rapidly with EMG velocity, see Eq.~(\ref{armature_melt_thickness}) and Figure~9. 

Finally, in Section V, we estimated the temperature rise of the armature, and the resulting thickness of the melted layer. The temperature rise in the armature may be a major limiting factor in EMG design, see Figure~9. 

In Section VI, we computed the efficiency of an EMG as a function of armature velocity and gun length. In this efficiency calculation, we only considered Joule heating,  kinetic energy of the armature, and the stored magnetic energy in the system, and we neglected all other energies, such as frictional heating between armature and rails.   We find that if the stationary part of the self inductance is small compared to the dynamic part of the self inductance (that of the rails), which is equivalent to long rails,  the efficiency of an EMG  increases with velocity.  Stated in another way, if the rails are long enough the EMG will have an efficiency that increases with velocity.  In all cases the upper limit of efficiency is 1/2 at high velocity.   However, with increasing velocity, the thickness of the melted layer in the rails also increases, see Figure~9.  The thickness of the melted layer also decreases with rail length.  Therefore, there is a tradeoff:  for sufficiently long rails, with increasing gun velocity the EMG is more efficient, however, at higher velocity the thickness of the melted layer is larger, which is likely to result in more damage to the rails.  Also, we may need rails that are too long for practical applications.  In our model of an EMG, we have neglected many practical effects, such as heating and ablation of armature/projectile at high velocities in the atmosphere.   

The most important conclusion is that, given the available choice of materials for rails and armatures, the gun has increased efficiency at higher velocity, however, there is increased  melting of the rails and damage is more likely  during each firing of high performance guns, where payload and range is designed for (high-velocity) naval guns.  This analysis was performed for simple rail guns. Other geometries exist, although all appear to rely on the sliding armature to carry the high currents and hence will generate a characteristic skin effect that results in significant rail heating. 
 
Finally, heating of the rails on a microscopic scale depends on inhomogeneities of the metal, such as crystalline alignment and dislocations.  These inhomogeneities will lead to spatial fluctuations of the current density. This will likely lead to a spatially inhomogeneous deposit of energy, and local ``hot spots".  Such spatial fluctuations in the case of an EMG are likely to lead to the analogue of what is usually termed ``gun barrel erosion".  A comment made at the end of the chapter written by I. Ahmad in 1988,  entitled ``The Problem of Gun Barrel Erosion: An Overview", is still  highly relevant: ``Other forms of erosion/corrosion problems might appear as a result of advances made in the development of liquid propellant guns and electromagnetic gun technologies. In fact, the erosion in the latter could be quite severe, as it involves interaction of high-temperature plasma with the launch surface causing it to partially melt at each firing event. It will be necessary to identify materials and design methodologies to minimize these problems."  Our paper does not include any effect of the high-temperature plasma, but it shows the effect of Joule heating and geometric design considerations on the rails.


%
%


\section*{Acknowledgment}
This work was sponsored in part by ILIR at the AMRDEC.


\begin{thebibliography}{1}
\bibitem{explain}The first part of this assumption is conservative, since a deviation from 
${\frac{{\partial {\kern 1pt} I}}{{\partial \,x}}} $ increases the total heating, somewhat compensating for the optimistic assumption that 
$\frac{{\partial {\kern 1pt} I}}{{\partial \,y}} = \infty $  or $I=0$ at the boundary, see later discussion.
\end{thebibliography}
\end{document}